\documentclass[runningheads]{llncs}

\usepackage{graphicx}
\usepackage{placeins}
\usepackage{amsmath}
\usepackage{hyperref}
\usepackage{color}
\usepackage{xcolor}
\usepackage{booktabs}
\usepackage{multirow}

\newcommand{\df}{\delta\!f}
\newcommand{\CBV}{\textrm{CBV}}
\newcommand{\R}{\textrm{R}}
\DeclareMathOperator{\DicoBase}{Dico_{base}}

\DeclareMathOperator{\DicoVasc}{Dico_{vascular}}

\newcommand{\anonymisation}[1]{******}

\title{MARVEL: MR Fingerprinting with Additional micRoVascular Estimates using bidirectional LSTMs}
\titlerunning{MARVEL: MRF with Additional micRoVascular Estimates using biLSTMs}
\author{Antoine Barrier\inst{1}* 
\and
Thomas Coudert\inst{1}*
\and
Aurélien Delphin\inst{2}
\and
Benjamin Lemasson\inst{1}
\and
Thomas Christen\inst{1}
}

\authorrunning{A. Barrier, T. Coudert \emph{et al.}}
\institute{Univ. Grenoble Alpes, Inserm, U1216, Grenoble Institut Neurosciences, GIN, Grenoble, France  \\
\and
Univ. Grenoble Alpes, Inserm, US17, CNRS, UAR 3552, CHU Grenoble Alpes, IRMaGe, Grenoble, France}

\begin{document}
\maketitle              
\begin{abstract}

The Magnetic Resonance Fingerprinting (MRF) approach aims to estimate multiple MR or physiological parameters simultaneously with a single fast acquisition sequence. Most of the MRF studies proposed so far have used simple MR sequence types to measure relaxation times ($T_1$, $T_2$). In that case, deep learning algorithms have been successfully used to speed up the reconstruction process. In theory, the MRF concept could be used with a variety of other MR sequence types and should be able to provide more information about the tissue microstructures. Yet, increasing the complexity of the numerical models often leads to prohibited simulation times, and estimating multiple parameters from one sequence implies new dictionary dimensions whose sizes become too large for standard computers and DL architectures.
In this paper, we propose to analyze the MRF signal coming from a complex balance Steady-state free precession (bSSFP) type sequence to simultaneously estimate relaxometry maps ($T_1$, $T_2$), Field maps ($B_1$, $B_0$) as well as microvascular properties such as the local Cerebral Blood Volume (CBV) or the averaged vessel Radius (R).
To bypass the curse of dimensionality, we propose an efficient way to simulate the MR signal coming from numerical voxels containing realistic microvascular networks as well as a Bidirectional Long Short-Term Memory network used for the matching process.
On top of standard MRF maps, our results on 3 human volunteers suggest that our approach can quickly produce high-quality quantitative maps of microvascular parameters that are otherwise obtained using longer dedicated sequences and intravenous injection of a contrast agent. This approach could be used for the management of multiple pathologies and could be tuned to provide other types of microstructural information.

\end{abstract}

\section{Introduction}

The Magnetic Resonance Fingerprinting approach (MRF, \cite{Ma2013}) aims to estimate multiple MR or physiological parameters with a single fast acquisition sequence. 

The MRF process involves fast undersampled acquisitions with time-varying parameters that produce temporal signal evolutions (or fingerprints) in every voxel. These \emph{in vivo} fingerprints are then compared to a large number of simulated signals obtained using combinations of \emph{a priori} tissue parameters and stored in a ``dictionary'' database. The values of the parameters corresponding to the closest simulated signals or ``match'' are then assigned to the associated voxels, producing multiple quantitative maps simultaneously. 

Most of the MRF studies proposed so far have used simple MR sequence types, such as spoiled gradient echo, for their fingerprints and have focused on the measurements of the transverse and longitudinal relaxation times $T_1$ and $T_2$ as well as the transmit field $B_1$ \cite{jiang2015mr,ma2017slice}. Even with this small number of dimensions in the dictionary, multiple strategies had to be proposed to reduce the long matching times initially obtained using direct dot product analysis ($>$ hours) as well as the large sizes of the dictionaries ($>10$\,Gb). This includes data compression with SVD decomposition \cite{MA_SVD}, fast group matching \cite{cauley2015fast}, and various deep learning architectures for the matching step including dense structures \cite{Cohen2018}, convolutional \cite{Fang2017,Hoppe2018} and recurrent \cite{Hoppe2019,Cabini2024} networks or auto-encoders \cite{Golbabaee2021}. 

In theory, the MRF concept could be used with a variety of other MR sequence types and should be able to provide more information about the tissue microstructures. As long as the fingerprints are made sensitive to the parameters of interest, are different from each other and the simulations are realistic enough to capture the physical processes of interest. However, increasing the complexity of numerical models often leads to prohibited simulation times. Similarly, estimating multiple parameters from one sequence implies new dictionary dimensions whose sizes become too large for standard computers and that even standard deep learning architectures have not been able to handle well. 

In this paper, we propose to analyze the MRF signal coming from a complex balance Steady-state free precession (bSSFP, \cite{Scheffler2003PrinciplesAA}) type sequence that is known to be sensitive to various biological parameters including vascular microstructures. Our goal is to simultaneously estimate relaxometry maps ($T_1$, $T_2$), Field maps ($B_1$, $B_0$ or corresponding frequency shift $\df$) as well as microvascular properties such as the local cerebral blood volume (CBV) or the averaged vessel Radius (R). These latter vascular properties are of interest for the management of multiple pathologies including stroke or cancer but are usually acquired with much longer dedicated MR sequences and require intravenous injection of a contrast agent. 

\subsubsection{Contributions. }
In order to bypass the curse of dimensionality, we propose:
\begin{enumerate}
    \item An efficient way to simulate MR signals coming from numerical voxels containing realistic microvascular networks. Inspired by Wang et al. \cite{Wang2019}, we first estimate the frequency ($\df$) distributions inside voxels and convolve standard MRF dictionaries along this dimension. In this way, only small dictionaries have to be stored and fast vascular simulations can be made on demand.
    \item A Bidirectional Long Short-Term Memory (BiLSTM, \cite{GRAVES2005602}) network was used for the matching process. These types of networks have shown promise in modeling sequential data, making them suitable for analyzing MRF temporal sequences \cite{SIAMI}. The bidirectionality is used to further improve the sensitivity to all parts of the fingerprints and provide accurate measurements for the 6 parameters of interest $T_1$, $T_2$, $B_1$, $\df$, CBV and R. The network is trained with on-fly simulations to avoid storing the entire dictionary.
\end{enumerate}

\section{Material and Methods}

This section describes the material and methods used for our experiments. The associated code can be found at \url{https://github.com/nifm-gin/MARVEL}.

    \subsection{Towards Vascular Dictionaries of Signals}

The signal response of a voxel to a bSSFP-type sequence is impacted by its underlying microvascular properties. We explain how to extend standard dictionaries based on Bloch equations to take into account these intravoxel structures.

        \subsubsection{Base Dictionary Generation using Bloch Equations. }
The Bloch equations given below describe the evolution of the nuclear magnetization vector $\mathbf{M} = (M_x, M_y, M_z)$ as a function of the longitudinal and transverse relaxation times $T_1$ and $T_2$, the surrounding magnetic field vector $\mathbf{B}_0 = (B_x, B_y, B_z)$ with corresponding frequency shifts $\df$ and the gyromagnetic ratio $\gamma$: 
{\scriptsize
\begin{align*}
\scriptsize
\frac{dM_x}{dt} = \gamma\bigl(M_y B_z - M_z B_y\bigr) - \frac{M_x}{T_2} \,, &\qquad
\frac{dM_y}{dt} = \gamma\bigl(M_z B_x - M_x B_z\bigr) - \frac{M_y}{T_2} \,, \\
\frac{dM_z}{dt} = \gamma\bigl(M_x B_y &- M_y B_x\bigr) - \frac{M_z - M_0}{T_1} \,, 
\end{align*}
\normalsize
where $M_0$ is the steady-state nuclear magnetization. 
Numerical simulations of those equations allow to compute a $4$-dimensional dictionary, called $\DicoBase$, with signal evolutions associated to a set of provided tissue parameters $T_1$, $T_2$, $B_1$ and $\df$. To generate this dictionary, we used Python combined with a Matlab code derived from a reference Bloch simulator \cite{Bloch_Simulator} for standard relaxometry sequences. Simulations were performed using a main magnetic field of 3\,T.
}

        \subsubsection{Vascular Dictionary Generation. }
\begin{figure}[!t]
    \includegraphics[height=8.5cm]{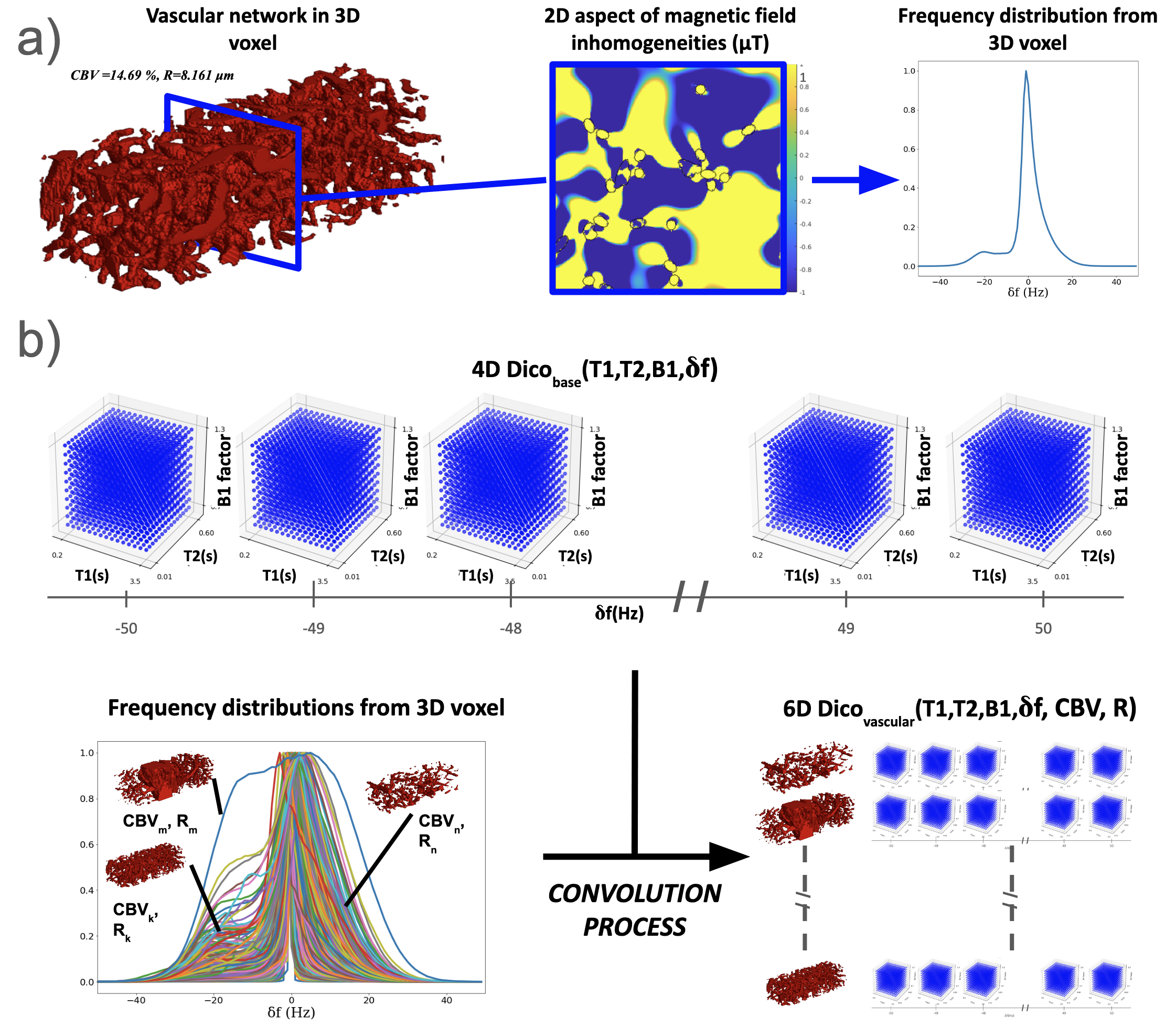}
    \centering
    \caption{a) Simulations of intra-voxel frequency distribution. b) Creation of a vascular MRF dictionary using a 4-dimensional base dictionary and distributions of frequencies.}
    \label{fig:inhomogeneities}
\end{figure}

\emph{In vivo}, there is usually more than one $\df$ value inside the voxels. This comes from the interaction of the subvoxel microstructures with different magnetic susceptibilities (such as blood vessels) with the main magnetic field of the scanner. It is possible to simulate the magnetic field spatial distribution created by realistic vascular networks inside 3D voxels (see Delphin et al. \cite{Delphin2023} for example) and in turn provide realistic distributions of $\df$ values inside each voxel, as illustrated in Figure~\ref{fig:inhomogeneities} a). These $\df$ distributions depend on the CBV and R values characterizing the vascular network. In order to generate a vascular dictionary, composed of signal evolutions that take into account the influence of CBV and R, we convolve the Bloch dictionary $\DicoBase$ with $\df$ distributions estimated from 3D vascular structures. The convolution process is illustrated in Figure~\ref{fig:inhomogeneities} b). Eventually, a dictionary $\DicoVasc$ with 6 MRI parameters $T_1$, $T_2$, $B_1$, $\df$, CBV and R can be built (but it is also possible to compute the vascular dimensions without having to store the full 6-dimensional dictionary). Note that a similar process has been used by Wang et al. \cite{Wang2019} to estimate the $T_2^*$ relaxation times using Lorentzian $\df$ distributions. 

    \subsection{Standard Dictionary Matching Process}

As a reference, a dictionary-matching process was used to provide quantitative parameter maps. Each voxel signal of the acquisition is matched to the signal of $\DicoVasc$ that maximizes the inner product, allowing for retrieval of the associated tissue parameters. This standard matching approach faces several limits when considering a large number of MRI parameters, mainly due to dictionary storage constraints, computation time issues, and the small number of values per parameter that can be simulated. For our reconstructions, a base dictionary $\DicoBase$ containing 43,000 signals was computed for a range of 10 $T_1$ values (from 0.2 to 3.5\,s), 10 $T_2$ values (7 values from 10 to 200\,ms and for 200, 400 and 600\,ms), 5 $B_1$ values (from 0.7 to 1.2) and frequency offset $\df$ values (from -50 to 49\,Hz with an increment of 1\,Hz), keeping only signals for which $T_1 > T_2$. Then, a vascular dictionary $\DicoVasc$ was generated by convolving the 43,000 entries of $\DicoBase$ with 300 $\df$ distributions coming from vascular structures obtained from 3D microscopy imaging (see \cite{Delphin2023}). To ensure fully defined distributions during the convolution process, only $\df$ values between -30 and 30\,Hz were used in the expanded dictionary. Even with a small number of vascular $\df$ distributions, the final dictionary contains 7,344,400 entries and the file size is already 29.4\,Gb.

    \subsection{Deep Learning Reconstruction}

We introduce a deep learning reconstruction framework in order to overcome the limits of dictionary-matching in high dimension and allow the computation of quantitative parameter maps in a reasonable time for clinical applications.

        \subsubsection*{A Bidirectional Recurrent Network. }
Observing that unidirectional LSTM structures \cite{Hoppe2019,Cabini2024} did not seem suited for the difficulty of the task (as shown in Figure~\ref{fig:cartesian_reco_comparisons} for LSTM and Reversed LSTM, and discussed in Section~\ref{sec:results}), we decided to use a Bidirectional LSTM (BiLSTM) architecture which extends the capabilities of the network while preserving its simplicity: the bidirectional layer is followed by dense layers (additional details about the structure and parameters of the network are provided in the supplementary materials). We implemented the network in Python, using the TensorFlow library.

        \subsubsection*{Training \& Dictionary Generation. }
As explained in the previous section, the task of training a network to simultaneously estimate 6 MRI parameters requires a large number of microvascular distributions to learn the diversity of brain vascularization. To overcome the storage difficulties, we use a fixed base dictionary ($T_1$, $T_2$, $B_1$, $\df$) and compute, at regular steps of the training, a batched vascular dictionary of same size by convolving each signal of the base dictionary with a random microvascular frequency $\df$ distribution. The training procedure is detailed below. 

We generate the base dictionary with 1,000,000 signals associated to 10,000 triplets $(T_1, T_2, B_1)$, pseudo-randomly picked into $[0.2\,s, 3.5\,s]\times[0.01\,s, 0.6\,s]\times[0.7, 1.2]$ using a Sobol distribution, and to 100 $\df$ values (from -50 to 49\,Hz with an increment of 1\,Hz). The increasing number of signals in this dictionary used for convolution, compared to the matching case (with 43,000 entries), is made possible by the online update of the vascular expanded dictionary, which prevents the dictionary size from exploding when adding microvascular parameters. 

Before training, we compute a set of 28,000 voxel distributions of vascular parameters. Then, during training and at every 5 epochs, we generate a new training batched dictionary by randomly associating, to each set of parameters $T_1$, $T_2$, $B_1$, $\df$, a couple $(\CBV, \R)$ of vascular parameters among our 28,000 distributions, using the convolution procedure explained in the previous section. We only keep for training signals with $\df$ values between -30 and 30\,Hz (as in the matching case), leading to a total of 600,000 signals. Finally, to increase the robustness of the learning against noisy acquisitions, we add a centered Gaussian noise to signals of the vascular dictionary during training, with variance randomly chosen such that the resulting SNR is uniform in the range $[1, 20]$. This choice of a wide SNR range is motivated by previous studies \cite{Barbieri2021} and the aim of considering acquisitions with significant SNR variations (see next section). 


    \subsection{MRI \emph{in vivo} acquisition}

\emph{In vivo} acquisitions were realized on 3 healthy volunteers with a Philips 3\,T Achieva dStream MRI at the IRMaGe facility (MAP-IRMaGe protocol). The proposed MRF sequence was based on an IR-bSSFP acquisition. 260 repetitions were used (TR\,=\,21\,ms), with Flip Angle (FA) linearly increasing from 7° to 70° as suggested in \cite{Gomez}, and a quadratic phase cycle of 10°. To compare the robustness of the model against under-sampling noise, one acquisition was performed using Cartesian sampling (matrix size: $256\times256\times1$; voxel size: $0.78\times0.78\times3.00$\,mm$^3$) with a scan time of 12 minutes per slice, and the two other acquisitions were performed using a spiral trajectory (matrix size: $192\times192\times3$; voxel size: $1.04\times1.04\times3.00$\,mm$^3$) with 12 acquired shots and a scan time of  2 minutes per slice. The spiral trajectory enables faster scan times and is hence essential for the clinical use of MRF sequences. Yet, the $k$-space undersampling scheme used in spiral scanning induces significant noise on the acquired signals. A sequence with the same parameters scheme, except the application of a spoiling gradient in the slice selection direction, was also acquired in one volunteer (Spoil sequence). This sequence was used as a reference to show the ability of simple networks to reconstruct 3-dimensional MRF maps.

\section{Results}

\begin{figure}[p]
    \centering
    \includegraphics
    [width=\textwidth]{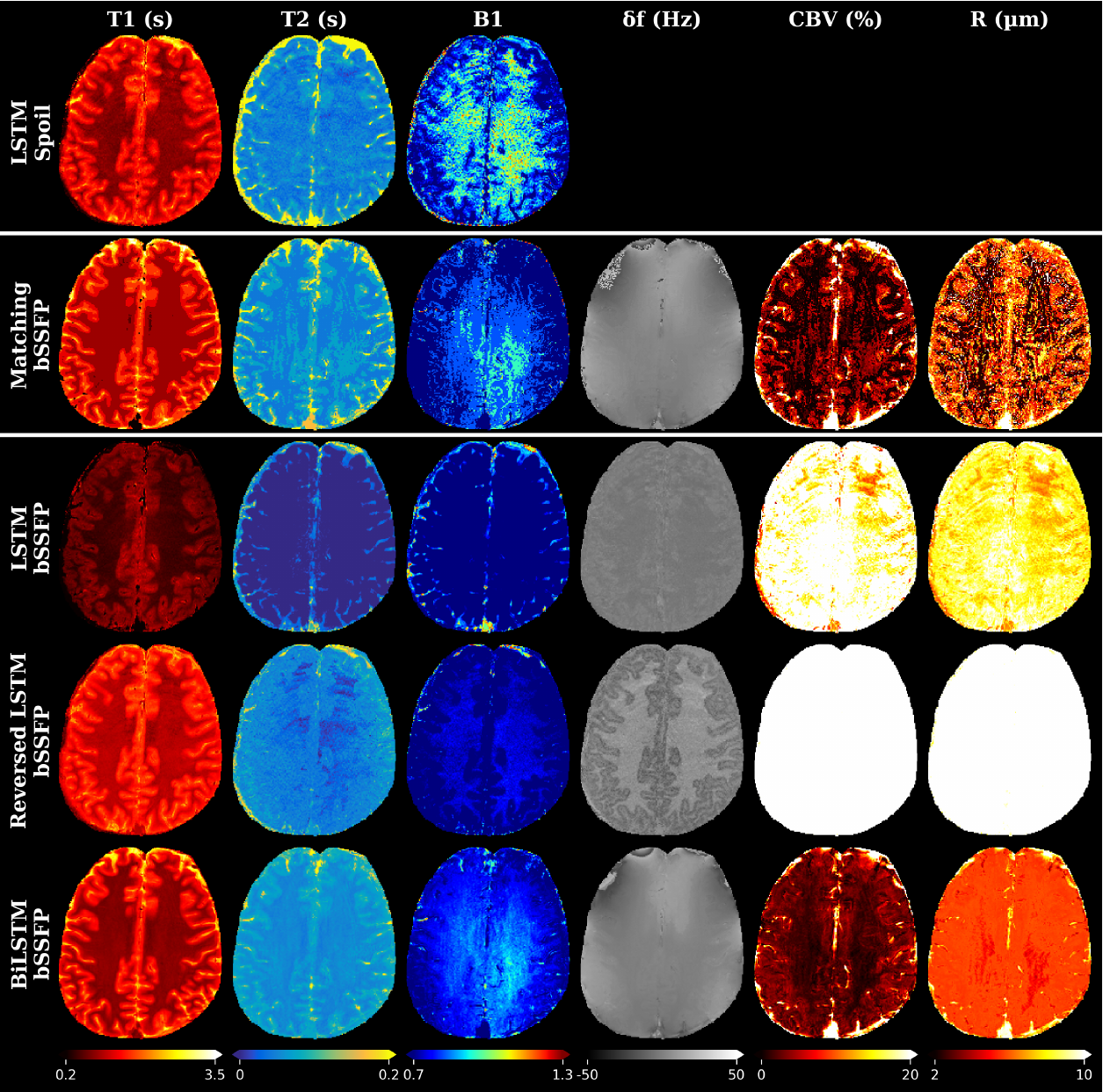}
    \caption{Parameter maps of the Cartesian acquisitions obtained with the reconstruction methods studied in this paper. (Note that the slice position slightly differs between the Spoil and bSSFP acquisitions.)}
    \label{fig:cartesian_reco_comparisons}
\end{figure}

\begin{table}[p]
\centering
  \caption{Mean and standard deviation of $T_1$, $T_2$, CBV and R reconstructed values in white matter (WM), grey matter (GM) and sagital sinus (manually drawn ROIs) in the slice of the Cartesian bSSFP sequence shown in Figure~\ref{fig:cartesian_reco_comparisons}. Best values compared to literature \cite{wansapura1999nmr,hasan2012human,gelman2001interregional,stikov2015accuracy,Delphin2023,ito2004database} are in \textbf{\textcolor{blue}{blue}}. }
  \label{tab:roi}
\scriptsize
\begin{tabular}{cc||c|c|c|c|c}
\bf\small Parameter & \bf\small Tissue & \bf\small LSTM & \bf\small Rev. LSTM & \bf\small BiLSTM & \bf\small Matching  & \bf\small Literature \\
 \hline \hline
 \multirow{2}[0]{*}{{\small$T_1$} (ms)} & WM & $538 \pm 121$ & $1119 \pm 177$  & \textcolor{blue}{$\mathbf{823 \pm 55}$} & $931 \pm 46$ & $	\sim690-1100$  \\
& GM & $674 \pm 202$  &  {$1440 \pm 261$} &   \textcolor{blue}{$\mathbf{1320 \pm 339}$} &   $1381 \pm 380$&  $	\sim1286-1393$ \\
\hline
 \multirow{2}[0]{*}{{\small$T_2$} (ms)} & WM & {$0.5 \pm 6$} &
 {$37 \pm 15$} & \textcolor{blue}{$\mathbf{54 \pm 5}$}  & {$50 \pm 13$} & {$	\sim56-80$} \\
 & GM &  {$8 \pm 22$} & {$53 \pm 21$} & {$69 \pm 21$}  &  \textcolor{blue}{$\mathbf{80 \pm 70}$} & $	\sim78-117$ \\
 \hline
 \multirow{3}[0]{*}{{\small CBV} (\%)} &WM  & $19.8 \pm 4.5$  & {$40.0 \pm 0.4$} & \textcolor{blue}{$\mathbf{2.0 \pm 0.9}$}  &  {$2.0 \pm 5.0$} &  {$\sim1.7-3.6$}\\
& GM & $22.2 \pm 5.3$ &  {$39.8 \pm 1.2$}  & \textcolor{blue}{$\mathbf{3.9 \pm 3.4}$}& {$1.49 \pm 1.9$} & $\sim3-8$\\
& Sag sinus & $19.5 \pm 8.5$ & {$37.3 \pm 4.2$}  & {$21.2 \pm 7.3$}& {$28.2 \pm 8.8$} & $ $\\
 \hline
 \multirow{3}[0]{*}{{\small R} ($\mu$m)} &WM  & $8.2 \pm 0.8$  &  {$10.0 \pm 0.0$} & \textcolor{blue}{$\mathbf{5.6 \pm 0.3}$}  & {$4.2 \pm 2.3$} &   {$6.8 \pm 0.3$}\\
& GM & $8.5 \pm 0.9$ &  {$10.0 \pm 0.0$}  &  \textcolor{blue}{$\mathbf{5.8 \pm 0.5}$}& {$5.4 \pm 2.2$} & $7.3 \pm 0.3$\\
& Sag sinus & $7.8 \pm 1.3$ & {$10.0 \pm 0.2$}  &  {$8.8 \pm 1.5$}& {$10.1 \pm 2.2$} & $ $
\end{tabular}
\end{table}

\begin{figure}[!t]
    \centering
    \includegraphics[height=3.5cm]{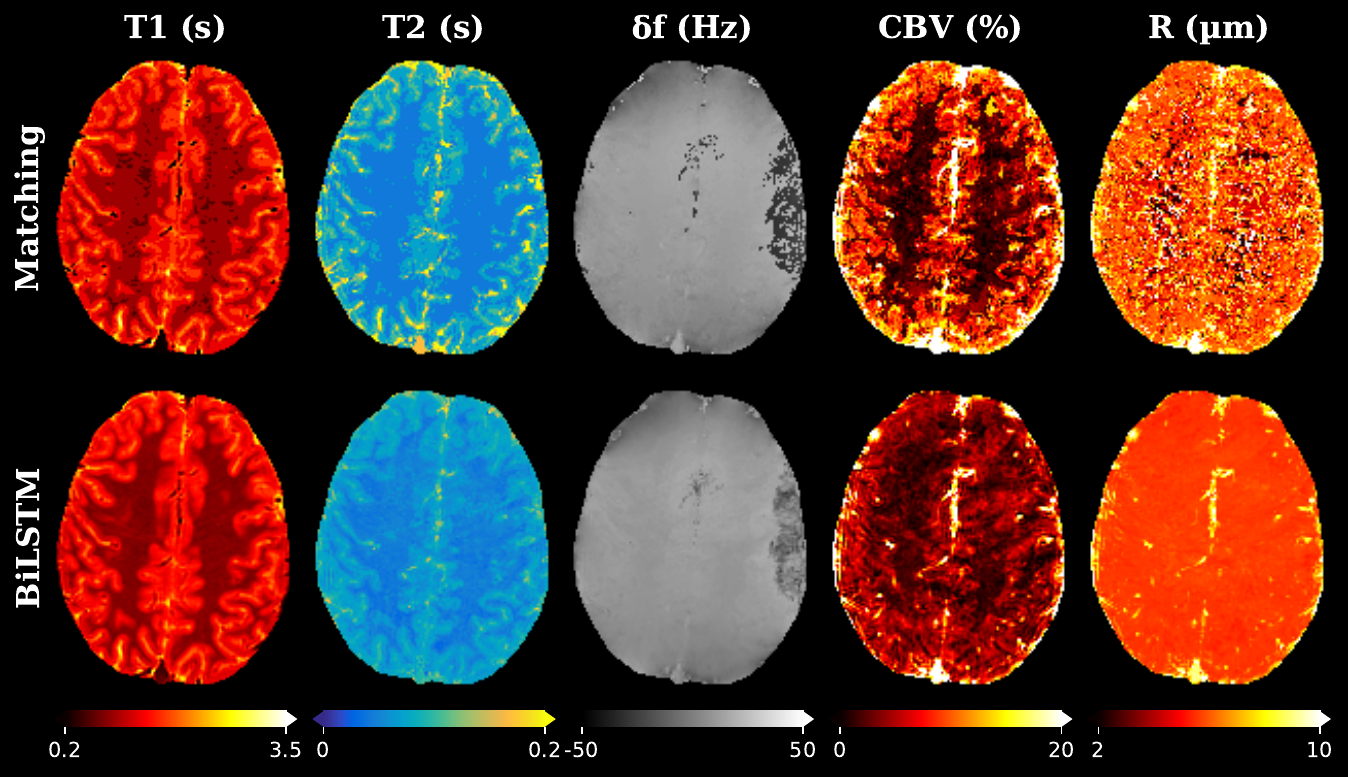}
    \hfill
    \includegraphics[height=3.5cm]{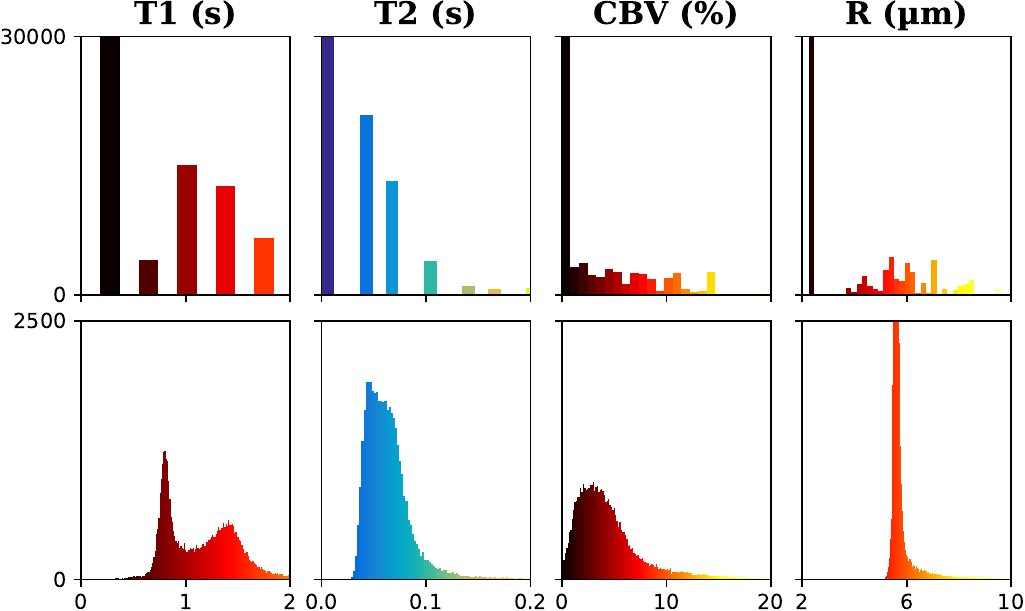}
    \caption{Parameter maps of one slice of a volunteer of the bSSFP spiral acquisition obtained with dictionary-matching and our {BiLSTM} network, and associated histograms. 
    }
    \label{fig:spiral_reco_comparison}
\end{figure}

Quantitative parameter maps obtained in one volunteer using Cartesian acquisition are given in Figure~\ref{fig:cartesian_reco_comparisons}. Relaxometry maps ($T_1$, $T_2$, $B_1$) obtained with the standard spoil sequence and reconstructed with the LSTM network are of high quality, suggesting that the network reconstruction works for low dimensional MRF acquisitions. Results from standard dictionary-matching on the bSSFP sequence (reconstruction time 2223\,s) show noisier maps but also provide frequency and microvascular maps with the right contrasts and global values in the expected ranges. This is not the case with the LSTM and the Reversed LSTM networks\footnote{For those networks, we replaced the bidirectional layer of our network by a unidirectional LSTM. The direction of input signals was inverted for the Reversed LSTM. } reconstructions. Our BiLSTM network is yet able to provide high-quality maps for all the parameters (reconstruction time 3.5\,s). In particular, the CBV and R maps are different from the relaxometry maps, with high values where large vessels are expected. This can also be observed in Table~\ref{tab:roi}, where BiLSTM values are closer to results obtained in previous literature studies\footnote{CBV and R values from the literature were obtained using contrast agent injection.}. Parametric maps obtained in one slice of the second volunteer using the spiral acquisition are presented in Figure~\ref{fig:spiral_reco_comparison}. Standard matching and BiLSTM reconstruction are compared and suggest that the network can adapt to a different type of space sampling and SNR. Histograms of values are also provided to highlight the discretized \emph{versus} continuous value distributions between the two approaches. BiLSTM results for all slices of the 2 volunteers using the spiral acquisition are given in the supplementary material. 

\section{Conclusion and Perspectives}   \label{sec:results}

In this study, we showed the possibility of quickly analyzing MRF data containing multiple dimensions including microvascular properties. This was done using a combination of fast and light realistic simulations and the use of Bidirectional LSTMs. Our first results on healthy volunteers are encouraging especially for the CBV maps that show a nice contrast between white matter, grey matter and blood vessels, and present similar values to those usually obtained using Gadolinium injections or TEP. Further analyses in multiple patients and comparisons with reference methods such as Dynamic Susceptibility Contrast (DSC) should still be conducted in order to validate the whole approach. 

Multiple solutions can already be foreseen to improve the results. The network structure could be further improved in order to handle longer fingerprints and the training step could contain additional undersampling noise in order to improve reconstruction from spiral acquisitions. Temporal compression methods such as Singular Value Decomposition \cite{MA_SVD} could also be applied to the MRF dictionary before training in order to help the LSTM networks capture long-term dependencies more effectively. 
In addition, the simulations could also be improved by using more realistic and diverse vascular vessel geometries and by taking into account other sources of magnetic susceptibility. Furthermore, the MRF sequence could be optimized with automatic procedures to improve the initial sensitivity to the parameters. 
Our approach has potential applications in the management of several pathologies including stroke and tumors. It might also be possible to extend the method to the measurement of other microstructural parameters such as brain oxygenation.

\begin{credits}
    \subsubsection*{\ackname} Project supported by the French National Research Agency [ANR-20-CE19-0030 MRFUSE].
We thank the MRI facility IRMaGe partly funded by French program ``Investissement d'avenir'' run by the French National Research Agency, grant ``Infrastructure d'avenir en Biologie et Santé'' [ANR-11-INBS-006].

    \subsubsection*{Disclosure of Interest.} The authors have no competing interests to declare that are relevant to the content of this article. 
\end{credits}

\FloatBarrier
\newpage

\bibliographystyle{splncs04}
\bibliography{biblio}

\newpage
\appendix
\title{\texorpdfstring{Supplementary Material for the paper \\ \bf MARVEL: MR Fingerprinting with Additional micRoVascular Estimates using bidirectional LSTMs}{Supplementary Material for the paper MARVEL: MR Fingerprinting with Additional micRoVascular Estimates using bidirectional LSTMs}}

\date{ }
\author{ }
\institute{ }
\maketitle

\begin{figure}[!ht]
\includegraphics[width=\textwidth]{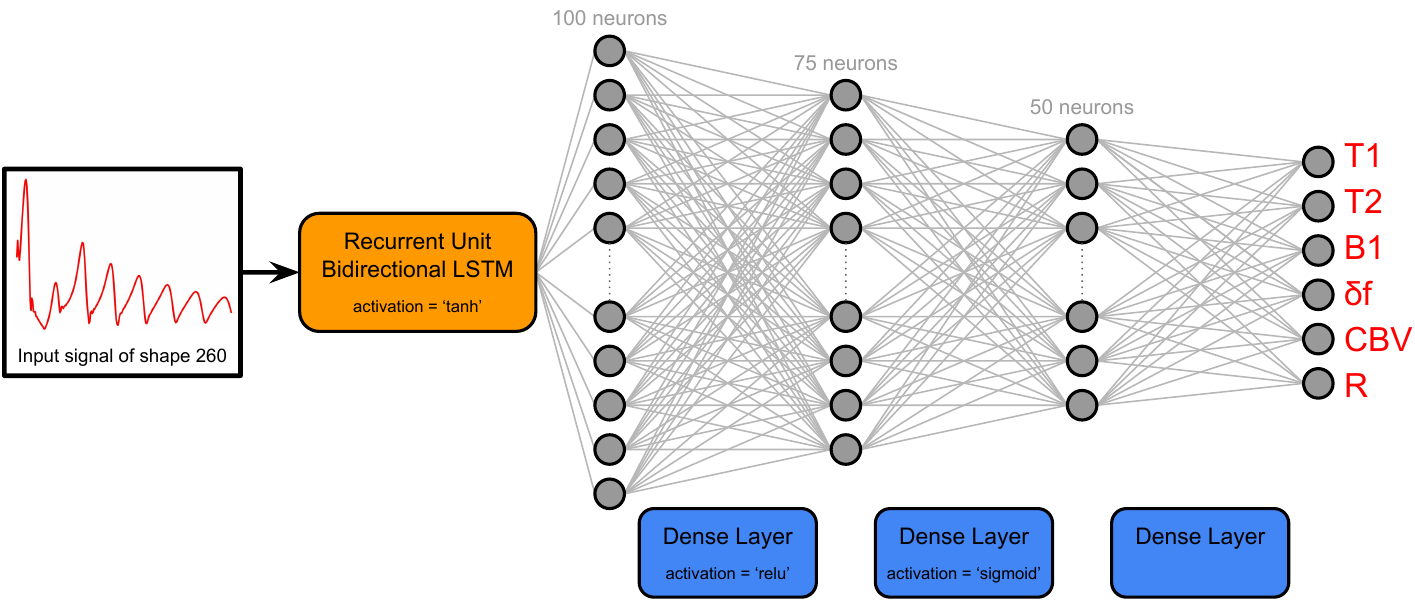}
    \caption{BiLSTM network structure. For training, we used the Adam optimizer with the MSE loss and an initial learning rate of $10^{-3}$, reduced by a factor of 0.8 every 5 epochs. }
    \label{fig:structure_NN}
\end{figure}

\begin{figure}[!ht]
    \centering
    \includegraphics[width=0.9\textwidth]{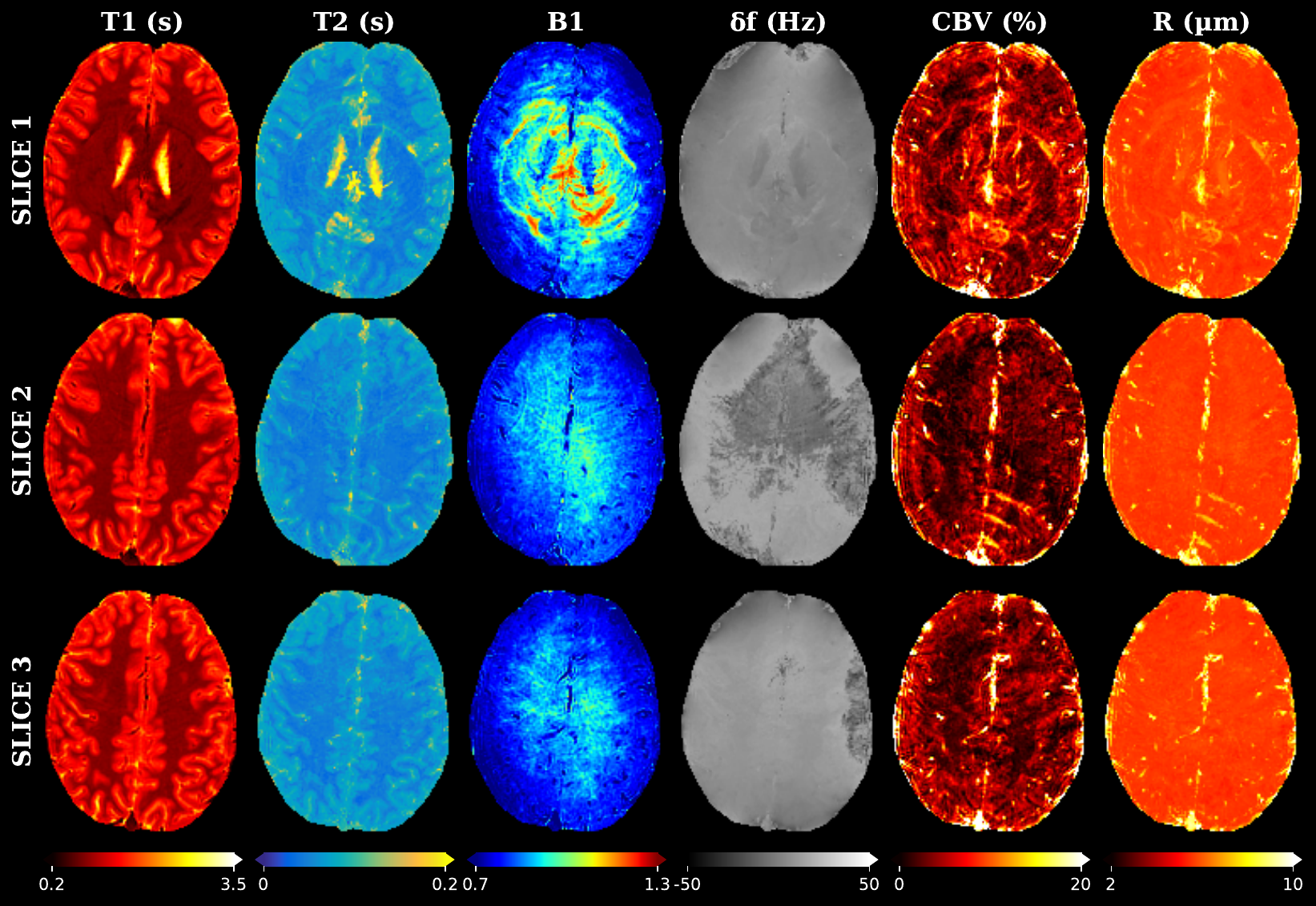}
    \caption{BiLSTM reconstructions of the spiral acquisition for subject 2. Computed in 3.5\,s. }
    \label{fig:spiral_reco_bilstm2}
\end{figure}   

\begin{figure}[!ht]
    \centering
    \includegraphics[width=0.9\textwidth]{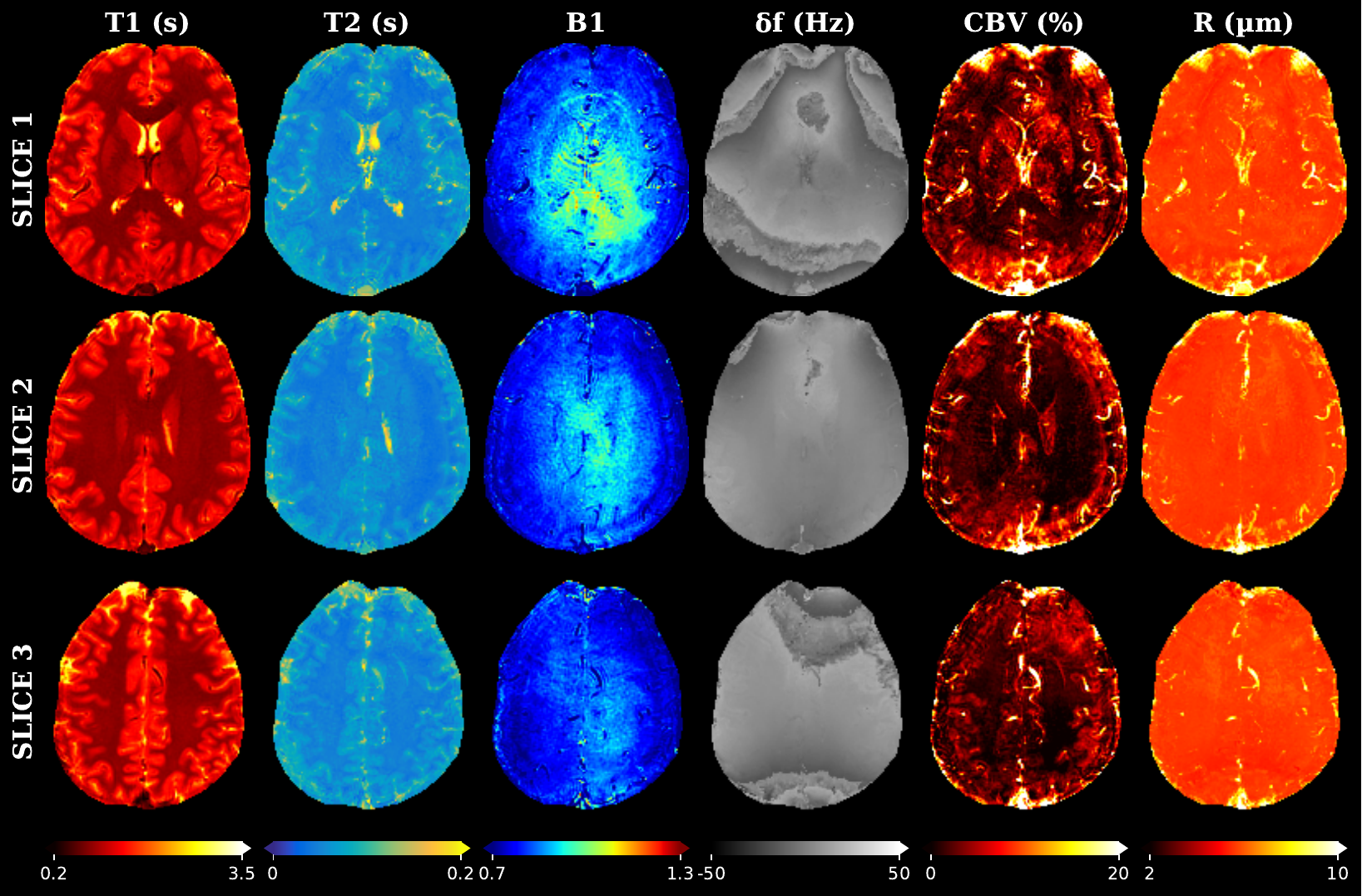}
    \caption{BiLSTM reconstructions of the spiral acquisition for subject 3. Computed in 3.5\,s. }
    \label{fig:spiral_reco_bilstm3}
\end{figure}

\end{document}